\documentclass[apj]{emulateapj}

\usepackage[]{hyperref}
\usepackage[greek,english]{babel}
\usepackage{longtable}
\usepackage{alltt} 
\usepackage{booktabs} 
\usepackage{multirow} 
\usepackage{overpic} 
\usepackage{xcolor}
\usepackage{seqsplit} 
\usepackage[caption=false]{subfig} 
\usepackage{units} 

\newcommand{\Sersic}{S\'{e}rsic}
\newcommand{\corser}{core-S\'{e}rsic}

\newcommand{\GALFIT}{\textsc{Galfit}}
\newcommand{\CORSAIR}{\textsc{Galfit-Corsair}}
\newcommand{\MBH}{$M_{\bullet}$}
\newcommand{\Msun}{$M_{\odot}$}
\newcommand{\magsb}{mag arcsec$^{-2}$}

\shorttitle{Questioning the biggest core ever reported for a galaxy (Holm~15A)}
\shortauthors{P. Bonfini, B. T. Dullo \& A. W. Graham}

\begin{document}

\title{Too big to be real? No depleted core in Holm~15A}
\shorttitle{Too big to be real? No depleted core in Holm~15A}

\author{
  Paolo Bonfini$^{\star}$, Bililign T. Dullo, and Alister W. Graham
}

\affil{Centre for Astrophysics and Supercomputing, Swinburne University of Technology\\ 
       Hawthorn, Victoria 3122, Australia}

\email{$^{\star}$pbonfini@swin.edu.au}

\begin{abstract}
 \noindent
 Partially depleted cores, as measured by \corser{} model ``break
 radii'', are typically tens to a few hundred parsecs in size.
 Here we investigate the unusually large ($R_{\gamma\prime = 0.5}$ = 4.57~kpc)
 depleted core recently reported for Holm~15A, the brightest cluster galaxy of
 Abell~85.
 We model the 1D light profile, and also the 2D image (using \CORSAIR{},
 a tool for fitting the \corser{} model in 2D).
 We find good agreement between the 1D and 2D analyses, with minor discrepancies
 attributable to intrinsic ellipticity gradients.
 We show that a simple \Sersic{} profile (with a low index $n$ and no depleted
 core) plus the known outer exponential ``halo'' provide a good description of
 the stellar distribution.
 We caution that while almost every galaxy light profile will have a radius
 where the negative logarithmic slope of the intensity profile $\gamma\prime$ 
 equals 0.5, this alone does not imply the presence of a partially depleted core
 within this radius.
\end{abstract}


\keywords{keyword: galaxies: elliptical and lenticular, cD --- galaxies: individual (\mbox{Holm~15A}) --- galaxies: photometry --- galaxies: structure}


\section[Introduction]{Introduction}
\label{Introduction}

\noindent
Luminous early-type galaxies ($M{_B}$ $\lesssim$ $-$20.5 $\pm$ 0.75~mag)
typically possess a core that is partially depleted of stars
\citep[e.g.][and references therein]{dullo:2014}.
This is evident by a marked flattening of their inner light distribution
\citep[][]{king:1966,king:1972,kormendy:1982,lauer:1983,byun:1996}.
The light profiles of these bright galaxies typically have a negative, inner
logarithmic slope $\lesssim$0.3, and have been fit using the King model
\citep{king:model_1,king:model_2}, the ``Nuker law''
\citep{grillmair:nuker,kormendy:1994,lauer:nuker,ravindranath,rest}, and
the \corser{} model \citep{graham:corser,trujillo:corser,ferrarese:2006}.
The Nuker model was designed to investigate the inner-most regions of
nearby early-type galaxies, and its double power-law nature was never
intended to adapt to the full, outer, intrinsically curved $R^{1/n}$\---{}like
profiles of galaxies \citep[e.g.][their Section 2]{faber}.
The outer $R^{1/n}$ curvature in galaxies results in a profile whose slope
changes as a function of radius.
As a consequence, the slope of a fitted Nuker model's outer power law
($\beta$) varies with the fitted radial extent of a galaxy.
Due to parameter coupling, it follows that all of the Nuker model parameters are
a function of the fitted radial extent \citep[][Figures 2--4]{graham:corser}.
The coupling is such that the Nuker model ``break radii'' are heavily
over-estimated, relative to the radius of maximum curvature in the actual profile,
i.e. the ``break'', and increasingly so as the fitted radial extent is increased
\citep{trujillo:corser}.
This contributed to \cite{lauer:2007} adopting the radius where the Nuker model
has a negative logarithmic slope of $\gamma\prime$ equal to 0.5 \citep{carollo}
as a measure of the core size.
All galaxy light profiles, even those with no depleted cores, have a radius
where $\gamma\prime$ equals 0.5.

In \cite{graham:corser} it was shown how a Nuker model (with an inner power-law
slope $<$0.3) can approximate a \Sersic{} profile (without any depleted core) if
the profile has a low \Sersic{} index and thus a shallow inner profile slope.
\cite{dullo:2013} showed that this has occurred when modelling several real
galaxies.
Most recently, for example, \cite{krajnovic:2013} report that their ATLAS$^{3D}$
galaxy NGC~4473 is a core galaxy according to their Nuker model fit.
However, it actually contains an additional nuclear component rather than a
depleted core \citep{krajnovic:2006}, which \cite{dullo:2014} have shown through
the use of the \Sersic{} and \corser{} models.
The core-Sersic model was introduced, in part, to prevent confusion when a profile
has a shallow inner slope but no central deficit of stars, and to
provide more robust, physically meaningful radii, slopes, and flux deficits
for partially depleted cores. 
   
Core sizes, as measured by the break radii of the \corser{} model ($R_{b}$), are
typically tens to a few hundred parsecs
\citep[e.g.][]{trujillo:corser,ferrarese:2006,richings,dullo:2012,dullo:2013,dullo:2014,rusli}.
These \corser{} break radii are where the \corser{} model has its maximum
curvature, and, for galaxies with depleted cores,
\citet[their Section 5.2]{dullo:2012} revealed that this also matches well with
the cusp radius\footnote{
 The cusp radius $R_{\gamma\prime}$ is defined as the radius at which the negative
 logarithmic slope of the intensity profile $\gamma\prime$ equals a pre-specified
 value \citep[][]{carollo}.
 Hereafter we will use $R_{\gamma}$ to actually indicate $R_{\gamma\prime=0.5}$.
 \label{footnote:r_gamma}
}
$R_{\gamma\prime}$ where the negative logarithmic slope of the projected light
profile equals 0.5.
As noted, galaxies without partially depleted cores can also possess such a
radius, and therefore one still needs to establish if there is an inner deficit
relative to the outer profile.
Indeed, many galaxies have an inner profile slope shallower than 0.5, and 0.3,
but do not have a partially depleted core
\citep[][their appendix A.2]{graham:2003,dullo:2013}.
A thorough review of galaxy light profiles can be found in \cite{graham:review_profiles}.

One of the most accredited scenarios for core formation attributes the depletion
to the scouring action of black hole (BH) binaries formed during dry galaxy
merger events.
The binary depletes the centre of a galaxy by ejecting stars via three-body
interactions which result in the orbital decay of the binary
\citep[e.g.][]{begelman,ebisuzaki,milosavljevic,merritt}.

\cite{merritt} suggested that the stellar mass depleted via the binary BH
scouring mechanism scales as 0.5$N$\MBH{}, where $N$ is the effective number of
major dry mergers which the galaxy experienced.
However, the observational result that $M_{def}$ is typically 0.5--4 \MBH{}
\citep[e.g.][]{graham:2004,ferrarese:2006,hyde,rusli,dullo:2014},
implies that the core galaxies characterized by the larger $M_{def}$/\MBH{}
ratio should have experienced up to 8 major dry mergers.
Such a number of major mergers near or above this figure is excessive if compared
to the merger rates derived from the observation of close massive galaxy pairs in
the local ($z$ $<$ 0.8) Universe \citep[e.g.][]{bell,depropris,casteels}.
A second mistery is that \cite{savorgnan:2015} have revealed that the large bulges
with over-massive BHs (or low velocity dispersions) in the \MBH{}--$\sigma$
diagram do not have larger $M_{def}$/\MBH{} ratios as expected from dry mergers and
as suggested by \cite{volontieri}. 

The depleted core radius is indicative of the central mass depletion
experienced by the galaxy, modulo the pre-existing central stellar density
profile.
Scaling relations exist between the final merged mass (\MBH{}) of the central
super-massive black hole (SMBH) and both the ejected stellar mass
\citep[e.g.][]{graham:2004,ferrarese:2006} and the core radii
\citep[e.g.][]{lauer:2007,dullo:2013,dullo:2014,rusli}.
These studies have described the \MBH{}--(core radius) correlation using
a log-linear relation, albeit with large uncertainties due to significant
scatter in the data.
The ongoing effort to characterize the high mass end of the relations
is necessary to investigate whether a single slope is appropriate to describe the
\MBH{}--$R_{b}$ diagram, as is currently assumed in the range
10$^{8}$ $\lesssim$ \MBH{}/\Msun{} $\lesssim$ 10$^{10}$.
A bend in the \MBH{}--$R_{b}$ relation may, for example, indicate different regimes
of efficiency for the BH scouring mechanism.
For instance, \cite{kulkarni} have suggested that multiple (i.e., more than 2)
SMBH systems can significantly increase the effectiveness of the scouring activity,
and generate cores with a mass deficit up to five times \MBH{}. 

A non-linear relation may also indicate that additional mechanisms other
than binary BH scouring are operating.
For example, a ``kicked'' SMBH can create an enhanced depleted core by crossing
the nucleus multiple times \citep[e.g.][]{redmount,merritt:2004,boylan,gualandris}.
In this scenario, the SMBH is placed on a radial orbit, intersecting with the nucleus,
after the recoil acquired upon its creation from the coalescence of a
specially-oriented BH binary.
Such a recoil is generated in response to the linear momentum carried away by the
anisotropic emission of gravitational waves.

Another possibility for creating large cores is  the ``stalled binary'' model
\citep{goerdt}, which proposes that the scouring activity is performed by a captured
in-falling object.
In this picture, a ``perturber'' spirals towards a galaxian center
due to dynamical friction.
The ``stalled binary'' scenario of \cite{goerdt} predicts core radii up to $\sim$3~kpc
and $M_{def}$ scaling as the mass of the perturber.
In doing so, it exerts a tidal action on the central mass distribution, shredding
it and creating a partially depleted core.
This model predicts that the core radius will correspond to the orbit at which the
infall of the perturber stalls due to a reduced efficiency of the dynamical friction.
In yet another scenario, \cite{martizzi} argued that the feedback action of
an active galactic nucleus (AGN) might be an important factor in the creation of cores.
The AGN feedback models by \cite{martizzi} easily produce core radii of sizes up to
$\sim$10~kpc although not yet confirmed.
In this model, AGN-driven gas outflows generate fluctuations in the gravitational
potential of the central region, from which stars are removed during the subsequent
re-virialization and adjustment process.
The expulsion of gas from the inner region might also induce an adiabatic expansion of
the central stellar distribution, hence flattening the central mass density profile.

Based on the extrapolation of the relation between the luminosity of the host
spheroid ($L$) and \MBH{} \citep[e.g.][]{graham:2013,graham:2015}, some Brightest
Cluster Galaxies (BCGs) are expected to host extremely massive BHs
(\MBH{} $\gtrsim$ 10$^{10}$~\Msun{}).
Recent N-body simulations including the effects of dark matter, black holes, as well
as baryons, have shown that black hole scouring in BCGs can create cores as large
as 3~kpc \citep{laporte}.
Therefore, the study of BCGs is important to explore the scarcely-populated
high-mass end of the \MBH{}--$R_{b}$ diagram, and hence provide better constraints
on the \MBH{}--$R_{b}$ scaling relation and the formation physics involved.

\subsection[The case of Holm 15A]{The case of Holm 15A}
\label{The case of Holm 15A}

\noindent
\citet[hereafter: LC14]{lopez} studied the BCG Holm~15A (D = 253~Mpc)\footnote{
 Luminosity distance from NED, relative to the Cosmic Microwave Background,
 assuming H$_{0}$ = 67.30~km s$^{-1}$ Mpc$^{-1}$, $\Omega_{m}$ = 0.315,
 $\Omega_{\Lambda}$ = 0.685
 \citep[\emph{Planck}+WMAP;][]{planck}.
 \label{distance}
},
located within the galaxy cluster Abell~85,
and reported the discovery of the largest depleted core known using the Nuker
model.
They found a cusp radius $R_{\gamma}$ = 4.57 $\pm$ 0.06~kpc,
which supersedes the record previously reported for the BCG in Abell~2261
\citep[$R_{\gamma}$ $\sim$ 3.2~kpc;][]{postman}.
The cusp radius of Holm~15A is not only more than three times larger than the average
$R_{\gamma}$ obtained by \cite{lauer:2007} using the Nuker model for their
sub-sample of $\sim$60 BCGs, but is also significantly larger than the biggest
recorded \corser{} model break radius ($R_{b}$ $\sim$ 1.5~kpc) reported by
\cite{hyde} for the massive elliptical galaxy SDSS J091944.2+562201.1.

LC14 derived $R_{\gamma}$ fitting a 2D Nuker model to an image of Holm~15A
within a major-axis radius of $\sim$80~kpc.
They confirmed their measurement of $R_{\gamma}$ non-parametrically from the 1D
radial light profile (i.e., applying the definition given in
Footnote \ref{footnote:r_gamma}).
However, their inner light deficit ($L_{def}$) was not calculated using the Nuker
model, but rather as the difference between a double-\Sersic{} fit (intended to
represent the actual galaxy light distribution) and a \cite{devaucouleurs} R$^{1/4}$
profile (intended to reproduce the ``pristine'' light profile prior to the
redistribution of the inner core light).
This unconventional light deficit was then used to derive a BH mass using
the $L_{def}$--\MBH{} relation of \cite{kormendy:2009}, giving an exceptionally
high \MBH{} $\sim$ 10$^{11}$~\Msun{}.
LC14 assessed the viability of this value by comparing it against the value of
\MBH{} estimated using other methods.
Since no direct (dynamical) measurement of the mass of the SMBH of Holm~15A
is available, LC14 resorted to using the scaling relations between \MBH{} and:
the stellar velocity dispersion ($\sigma$); the total luminosity of the
bulge; the Nuker model break radius; and $R_{\gamma}$
\citep[][and references therein]{kormendy:review}.
This analysis ultimately lead them to favor a more conservative
\MBH{} $\sim$ 10$^{10}$~\Msun{}.

\bigskip

Is the exceptionally large core of Holm~15A really due to a deficit of light relative
to the inward extrapolation of its outer light profile?
Is there an obvious and dramatic downward bend to its inner light profile?
We have further investigated the case of Holm~15A
by performing a detailed analysis of the light distribution of the galaxy.
For the first time, we apply the \corser{} model and check if a depleted core is
warranted over a core-\emph{less} model (i.e.\ a \Sersic{} model).

This paper is structured as follows.
In \S\ref{Data}, we present the data and the procedure used for modeling the
1D light profile and the 2D image of the galaxy.
In \S\ref{Discussion} we show that the light in Holm~15A is well fit by a
three-parameter \Sersic{} function plus an exponential ``halo'' and as such it does not
appear to have a partially depleted core with a well-defined break in the light profile.
We summarize our conclusions in \S\ref{Conclusions}.

\section[Data]{Data}
\label{Data}

\noindent
We have used an $r$-band\footnote{
 Nominally, in the $r$.MP9601 filter. This band corresponds to the
 SDSS $r$-band, to which we will refer hereafter.
 \label{footnote:filter}
}
image from the wide-field MegaPrime camera mounted on the
Canadian-French-Hawaiian-Telescope (CFHT).
The image was retrieved from the Canadian Astronomy Data Center\footnote{
 \href{http://www2.cadc-ccda.hia-iha.nrc-cnrc.gc.ca}{\seqsplit{http://www2.cadc-ccda.hia-iha.nrc-cnrc.gc.ca}}
} (CADC).
This enabled us to make a direct comparison with LC14, who used Kitt Peak National Observatory (KPNO) and CFHT data, in the $R$ and $r$ bands, respectively.
We have based our analysis on the CFHT data because --- to our knowledge --- they
represent the publicly available images for Holm~15A with the best seeing quality.
A narrow point spread function (PSF) is important to detect core structures,
whose projected sizes, even in the closest galaxies, are of the order of a few
arcseconds or less.

The detector of the CFHT MegaPrime camera is composed of 36 CCDs
(2112$\times$4644 pixels each) covering a sky area of $\sim$1 square degree.
Every MegaPrime image in the CADC is reduced via the Elixir pipeline \citep{elixir},
which performs the basic data calibration, including bias subtraction,
flat-fielding, de-fringing, and astrometric calibration.
The CFHT data used by LC14 was a single 120~s exposure.
We searched the CADC archive for a similar data set, and discovered that all the
images were severely affected by scattered light.
This scattered light, which appears as a diffuse radial gradient across the image,
is known to plague the MegaPrime data and it is due to unwanted internal reflections
through the optics.
This wavelength-dependent issue, especially prominent in the $r$-band, was
extensively reviewed by \cite{duc:2014}.

\begin{figure*}
 \makebox[\linewidth]{
  \begin{overpic}[width=0.48\textwidth]
   {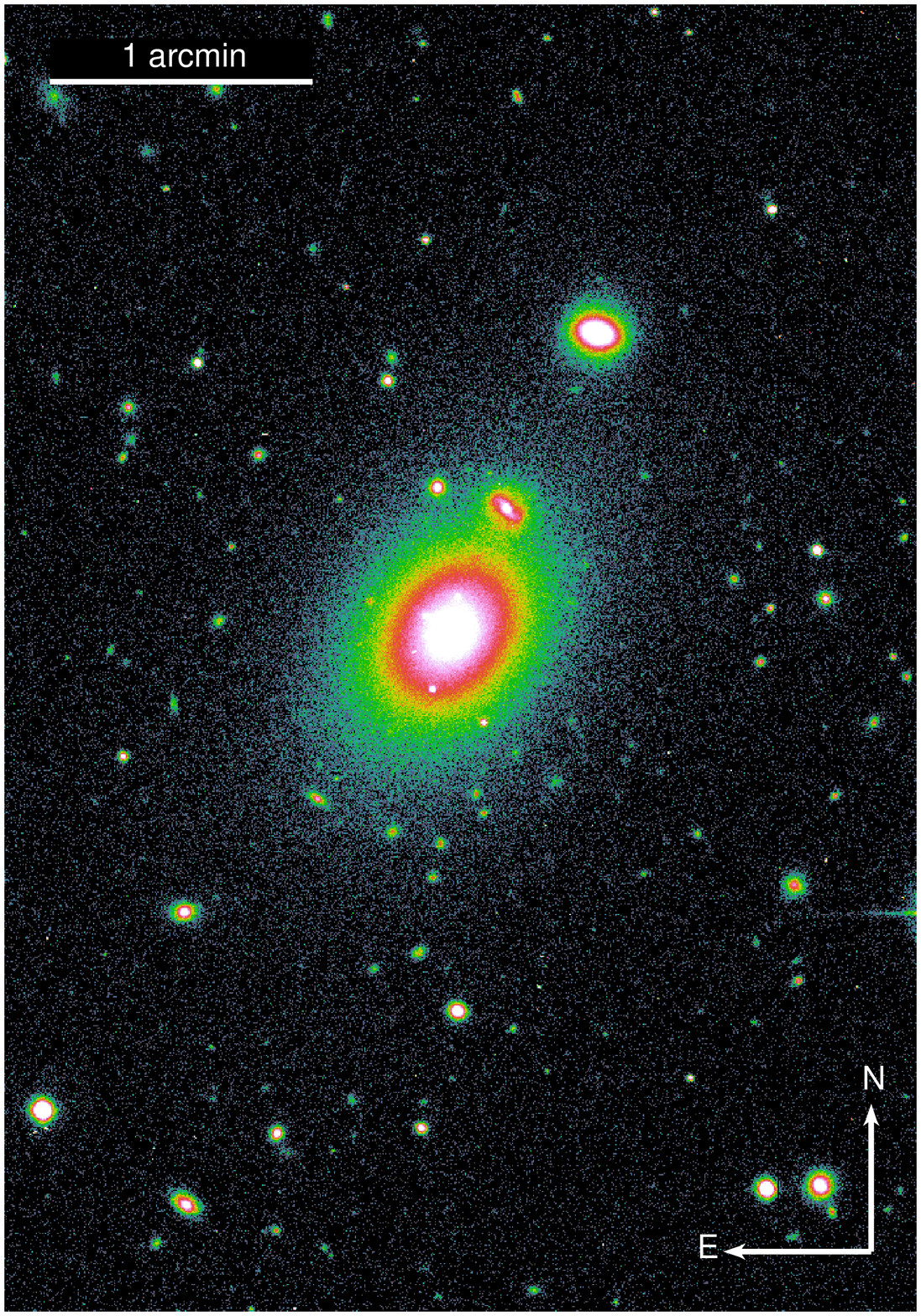}
   \put(5,5){\textcolor{white}{IMAGE}}
  \end{overpic}

  \begin{overpic}[width=0.48\textwidth]
   {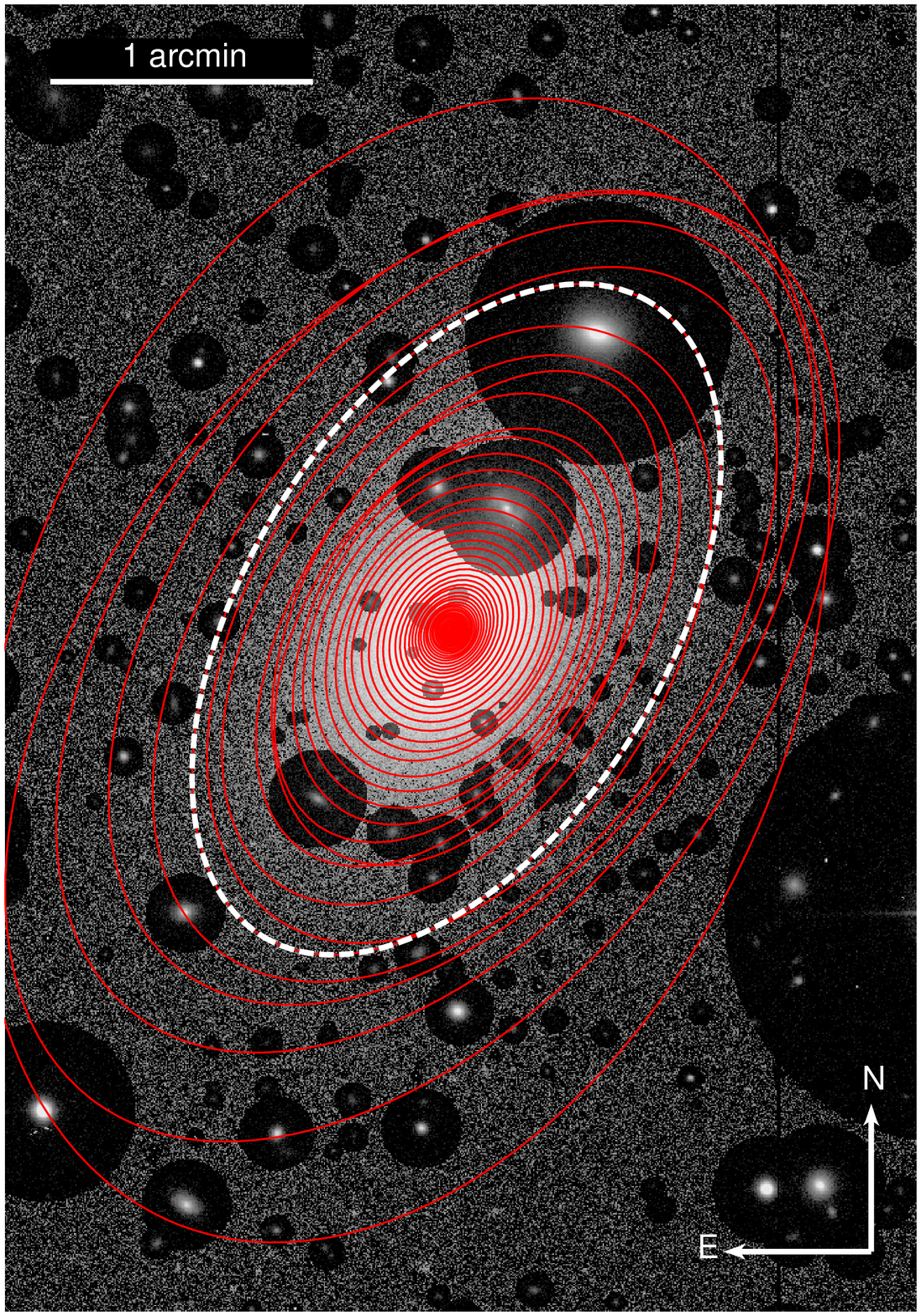}
   \put(5,5){\textcolor{white}{MASKED IMAGE}}
  \end{overpic}
 }
 \caption{
  CFHT-MegaPrime $r$-band image (\emph{left}) and relevant mask (\emph{right})
  for Holm~15A.
  \newline
  The masked areas (see \S\ref{Data}) have been arbitrarily decreased
  in intensity so as to still show the contaminating objects.
  We overplot the elliptical isophotes from IRAF.\emph{ellipse}
  (the boxiness of the ellipses is not considered in this representation).
  The dashed ellipse corresponds to the limiting surface brightness at which
  we truncated our 1D analysis ($\mu_{r}$ $\sim$ 25.5~\magsb{}; see
  \S\ref{Data}).
  This ellipse also corresponds to the physical extent of the 2D fit
  (everything outside the dashed curve was masked for the 2D fit;
  see \S\ref{Data}).
  \label{figure:mosaic_and_mask}
 }
\end{figure*}

Due to the lack of an analytical description for the scattered light, it could not
be readily disentangled from the smoother sky background.
We therefore modeled the combined background (sky + scattered light), and
subtracted the contribution from the two components at the same time.
Since Holm~15A easily fits within one single MegaPrime CCD, still allowing 
ample margins to determine the background, we performed this
operation and the other analyses only on the CCD which was centered on the galaxy.
First, we ran \textsc{SExtractor}  \citep{SExtractor} to determine the $R_{90}$
(i.e.\ the radius encompassing 90\% of the flux) of each source in the
image, and to obtain a median estimate of the remaining background flux and of the
background root-mean-square (RMS) fluctuation.
Each detected source was masked using an elliptical region with a major axis equal
to twice the size of the $R_{90}$ of the source, and with axis ratio
and position angle (P.A.) as determined by \textsc{SExtractor};
in addition, we masked any 5$\sigma$ fluctuation about the median background flux.
Finally, we used \GALFIT{} \citep{GALFIT} to fit a 2D gradient to the background,
adopting the \textsc{SExtractor} median background as the initial guess for the value
at the image center.
Figure \ref{figure:mosaic_and_mask} shows our background-subtracted $r$-band
image (see Table \ref{table:image} for image specifications).

\begin{deluxetable*}{cccccccc}
 \tabletypesize{\small}
 \tablecaption{CFHT-MegaPipe image characteristics \label{table:image}}
 \tablewidth{0pt}
 \tablehead{
  \colhead{Target}         &
  \colhead{RA  (J2000)}    &
  \colhead{Dec (J2000)}    &
  \colhead{$D$}            &
  \colhead{$m-M$}          &
  \colhead{Camera/Filter}  &
  \colhead{Exposure}       &
  \colhead{Scale} 
  \\
  \colhead{}                  &
  \colhead{[hh:mm:ss]}        &
  \colhead{[dd:mm:ss]}        &
  \colhead{[Mpc]}             &
  \colhead{[mag]}             &
  \colhead{}                  &
  \colhead{[sec]}             &
  \colhead{[$\arcsec$/pixel]} 
  \\
  \colhead{{\tiny (1)}} &
  \colhead{{\tiny (2)}} &     
  \colhead{{\tiny (3)}} &
  \colhead{{\tiny (4)}} &
  \colhead{{\tiny (5)}} &
  \colhead{{\tiny (6)}} &
  \colhead{{\tiny (7)}} &
  \colhead{{\tiny (8)}}
 }
 \startdata
Holm 15A & 00$^{h}$41$^{m}$50$^{s}$.5 & -09$\degr$18$\arcmin$11$\arcsec$ & 253 & 37.02 & MegaPrime/$r$ & 120 & 0.186 \\
 \enddata
 \tablecomments{
  Details of the CFHT-MegaPrime image used for the current work.
  \\
  $^{(1)}$ Target name.
  $^{(2,3)}$ Target coordinates from NED.
  $^{(4)}$ Luminosity distance from NED, corresponding to a redshift
           \mbox{$z$ $\sim$ 0.057} (see Footnote \ref{distance}).
  $^{(5)}$ Distance modulus.
  $^{(6)}$ CFHT camera and filter.
  $^{(7)}$ Total exposure time.
  $^{(8)}$ Image pixel scale.
 }
\end{deluxetable*}

\subsection[The 1D radial light profile]{The 1D radial light profile}
\label{The 1D radial light profile}

\noindent
An object mask was created using the \textsc{SExtractor} detections, and then
further refined by hand after visual inspection to exclude cosmic rays, hot pixels,
and smaller sources over-lapping with Holm~15A.
The right panel of Figure \ref{figure:mosaic_and_mask} shows the CFHT mosaic
with the masked areas down-scaled by an arbitrary amount; we also overplot the
elliptical isophotes derived using the IRAF.\emph{ellipse} task \citep{ellipse}.

\begin{figure*}
 \begin{center}
  \includegraphics[width=0.8\textwidth,angle=0]{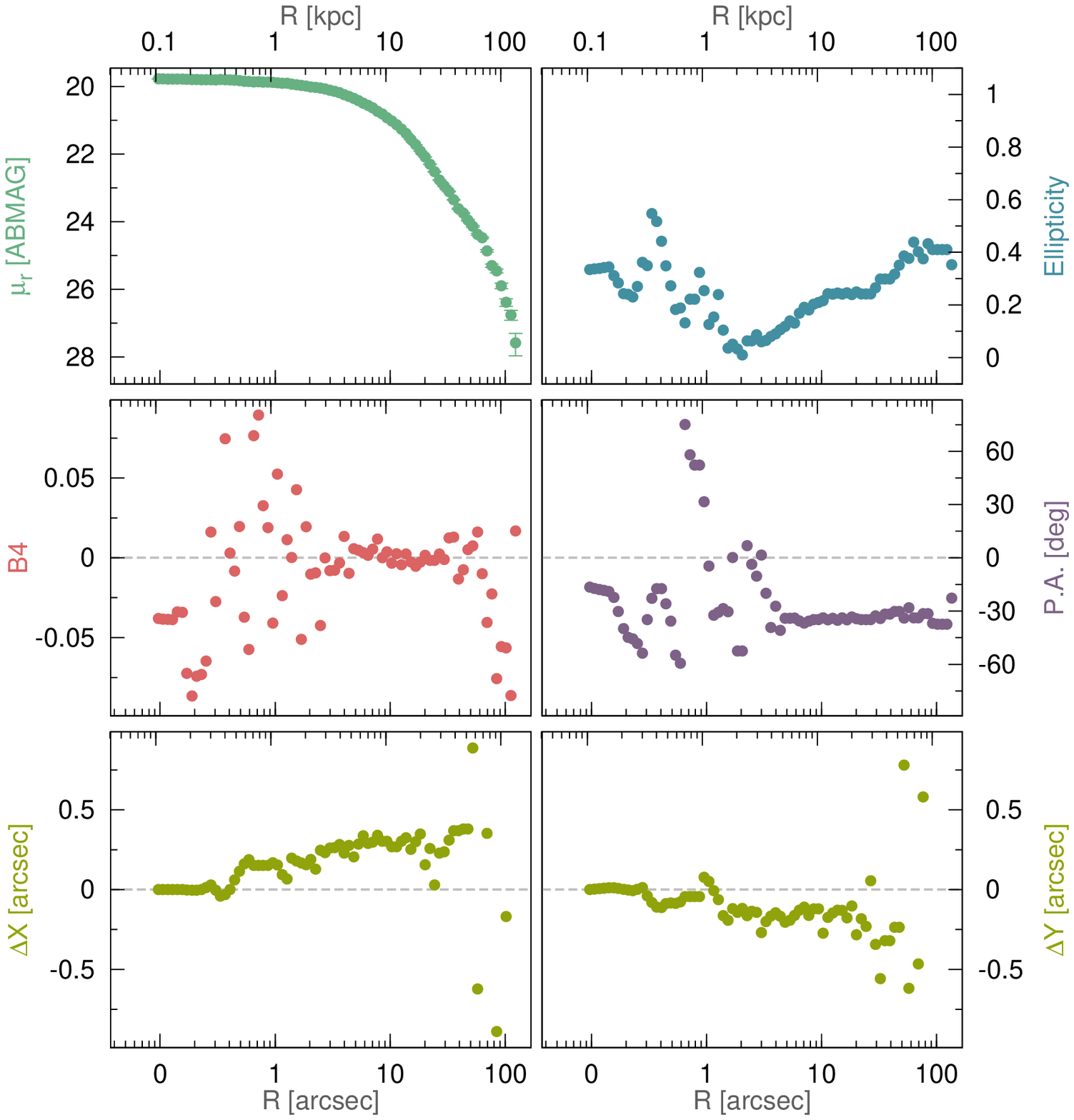}
  \caption{
   IRAF.\emph{ellipse} major-axis radial profiles for: $r$-band surface brightness
   (\emph{top-left}), ellipticity (\emph{top-right}), 4\emph{th} harmonic deviation
   from perfect ellipticity ($B4$; \emph{middle-left}), position angle
   (\emph{middle-right}), and isophote centroid shift along the $x$-axis
   (\emph{bottom-left}) and $y$-axis (\emph{bottom-right}) with respect to the
   innermost isophote.
   \label{figure:ellipse}
  }
 \end{center}
\end{figure*}

Figure \ref{figure:ellipse} shows the radial profiles along
the semi-major axis (SMA) obtained from IRAF.\emph{ellipse} for the: $r$-band
surface brightness ($\mu_{r}$); ellipticity ($e$); 4\emph{th} harmonic deviation
from perfect ellipticity ($B4$ "boxiness/diskiness" parameter); P.A.; and
isophote centroid shift.
We observe that beyond $\sim$2$\arcsec$, the ellipticity is steadily increasing,
except for a flattening in the range 10$\arcsec$--30$\arcsec$.
Increasing ellipticities have been shown to be common for BCGs, and they can be due
to the projection of the prolate (or triaxial) structure of the outer regions
(typically beyond $\sim$45~kpc) of these objects \citep[e.g.][]{porter}.

The fitting of models to the light profile (see \S\ref{The choice of models}) have been
performed via a Levenberg-Marquardt minimization procedure, as described in
\cite{dullo:2014}.
We chose to limit the fit to $\mu_{r}$ $<$ 25.5~\magsb{} (approximately
corresponding to SMA $<$ 90~kpc), based on where the residual large scale
background gradient (i.e., after the background subtraction process described above)
becomes comparable to the galaxy light gradient.
This limit roughly corresponds to the radius at which the elliptical isophotes stop
being concentric (see Figure \ref{figure:mosaic_and_mask}, right panel).

\subsection[The 2D image]{The 2D image}
\label{The 2D image}

\noindent
Modeling of the 2D light distribution was performed using
\CORSAIR{}\footnote{
 \href{www.astronomy.swin.edu.au/~pbonfini/galfit-corsair/}{\seqsplit{www.astronomy.swin.edu.au/{\textasciitilde}pbonfini/galfit-corsair/}}  
} \citep{corsair}, a tool developed to include the \corser{}
model into the \GALFIT{} fitting algorithm\footnote{
 \CORSAIR{} is retro-compatible with \GALFIT{}.
}.

To maximize the consistency with the 1D analysis, we restricted the fit within
the $\mu_{r}$ $\sim$ 25.5~\magsb{} isophote (see \S\ref{The 1D radial light profile})
by masking every pixel outside it (see Figure \ref{figure:mosaic_and_mask}, right
panel).
The \CORSAIR{} mask was hence obtained by combining this filter with the object
mask described before.
The PSF template was built with \textsc{PSFEx} \citep{psfex}, combining
$\sim$20 stars with signal-to-noise ratios ($S/N$) $>$ 100.
Since the variation of the PSF FWHM was negligible across the CCD (less than 5\%),
we selected stars from all areas.
The FWHM of the resulting PSF model is $\sim$0$\arcsec$.75, in agreement with
the value measured for real stars in the image.
Finally, the weight (``sigma'') image was constructed using the internal
\GALFIT{} algorithm, to which we supplied an estimate of the sky
RMS as measured on the image by using \textsc{SExtractor}.

We warn the reader that the varying ellipticity of Holm~15A
(Figure \ref{figure:ellipse}, top-right panel) can pose a challenge for the 2D
fitting, because in \CORSAIR{}, as with \GALFIT{}, each galaxy component is associated
with a single ellipticity and position angle.
When the components of real galaxies have ellipticity gradients, the model does not
allow for this.
That is, even when a galaxy is decomposed into multiple components, the problem might
persist whenever the ellipticity profile has a significant gradient over the range
where any single component dominates.
This issue is of no concern when modeling the 1D profile, which is simply
extracted along the major axis.
As such, small discrepancies are expected between the 1D and 2D analyses.

\subsection[The choice of models]{The choice of models}
\label{The choice of models}

\noindent
To investigate the presence of a partially depleted core in Holm~15A, we separately
fit a seeing-convolved \Sersic{} model and a seeing-convolved \corser{} model, in
both 1D and 2D.
As noted and modelled by LC14, Holm~15A has a halo of light around it.
The ``bump'' in the radial light profile at around 35~kpc,
which also corresponds to a step in the ellipticity profile
(see Figure \ref{figure:ellipse}), does indeed suggest that the galaxy hosts
a second component.
We therefore added an outer [seeing-convolved] exponential function to the
\Sersic{}/\corser{} ``bulge'', capturing the ``halo'' of Intra-Cluster Light (ICL).
An exponential halo model was used following \cite{seigar}, who revealed that
the halos of BCGs are typically exponential; see also \cite{pierini}.
We did additionally model the ICL with a \Sersic{} $R^{1/n}$ model, but we found
that it yielded a \Sersic{} index of $\sim$1, i.e. an exponential model.

The best-fit parameters obtained for the aforementioned models are reported
in Table \ref{table:fit}, while the model profiles are represented in Figure \ref{figure:fit_1D} (1D fit) and in the left and central panels of Figure
\ref{figure:fit_2D} (2D fit).
We stress that the profiles shown in Figure \ref{figure:fit_2D} represent a 1D
\emph{projection} of the 2D images and 2D models rather than what was actually
minimized, namely the difference between the 2D image and model, as shown in
lower panels of Figure \ref{figure:fit_2D}.
This projection was performed measuring the surface brightness of these images
along the isophotes identified with IRAF.\emph{ellipse} in our 1D analysis
(see \S\ref{The 1D radial light profile}).
In the next section, we will first mention the small discrepancies
between the results of the 1D and 2D modeling, and then proceed to discuss
the selection of the best-fit model.

\subsection[Fit comparison and choice of best-fit model]{Fit comparison and choice of best-fit model}
\label{Fit comparison and choice of best-fit model}

\begin{figure*}

 \begin{tabular}{p{0.55\textwidth} p{0.55\textwidth} p{0.05\textwidth}}

   \vspace{0pt}
   \begin{overpic}[angle=0,height=0.5\textwidth]
    {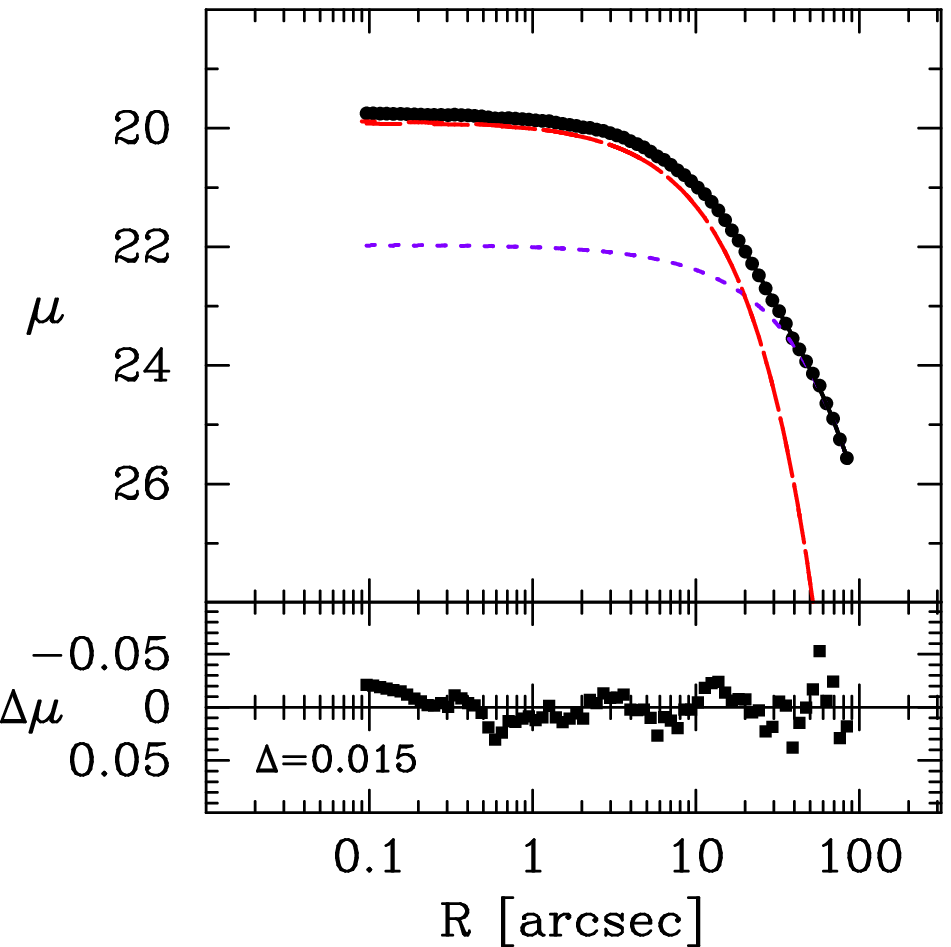}
    \put(28,40){\textcolor{black}{\textbf{\Sersic{}+exponential}}}
   \end{overpic}

   &
  
   \vspace{0pt}
   \hspace{-25pt}
   \begin{overpic}[angle=0,height=0.5\textwidth]
    {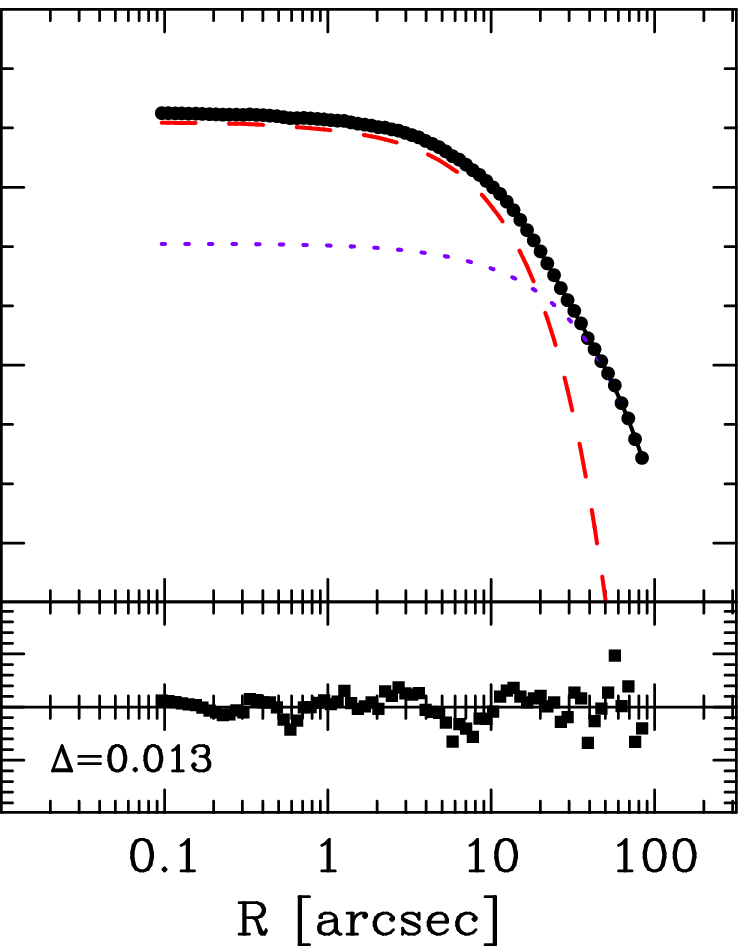}
    \put(8,41){\textcolor{black}{\textbf{\corser{}+exponential}}}
   \end{overpic}

 \end{tabular}
 
 \caption{
  1D light profile analysis:
  results of the \Sersic{}\-+exponential (\emph{left}), and  \corser{}\-+exponential
  (\emph{right}) model fit to the major-axis, $r$-band surface brightness profile
  of Holm~15A.
  \newline
  In the left panel, the red curve shows the \Sersic{} component.
  In the right panel, it represents the \Sersic{} portion of the \corser{}
  component.
  The exponential function which dominates at large radii is indicated by
  the violet short-dashed curves.
  The solid curve represents the complete fit to the profiles, with the RMS
  residuals, $\Delta$, about each fit given in the lower
  panels.
  Note that the core part of the \corser{} model fits an apparent slight
  ``excess'' of light over the inner 0$\arcsec$.5, rather than a light ``deficit''.
  \label{figure:fit_1D}
 }
\end{figure*}

\begin{figure*}

 \begin{tabular}[t]{c}

  \begin{overpic}[width=1.0\textwidth]
   {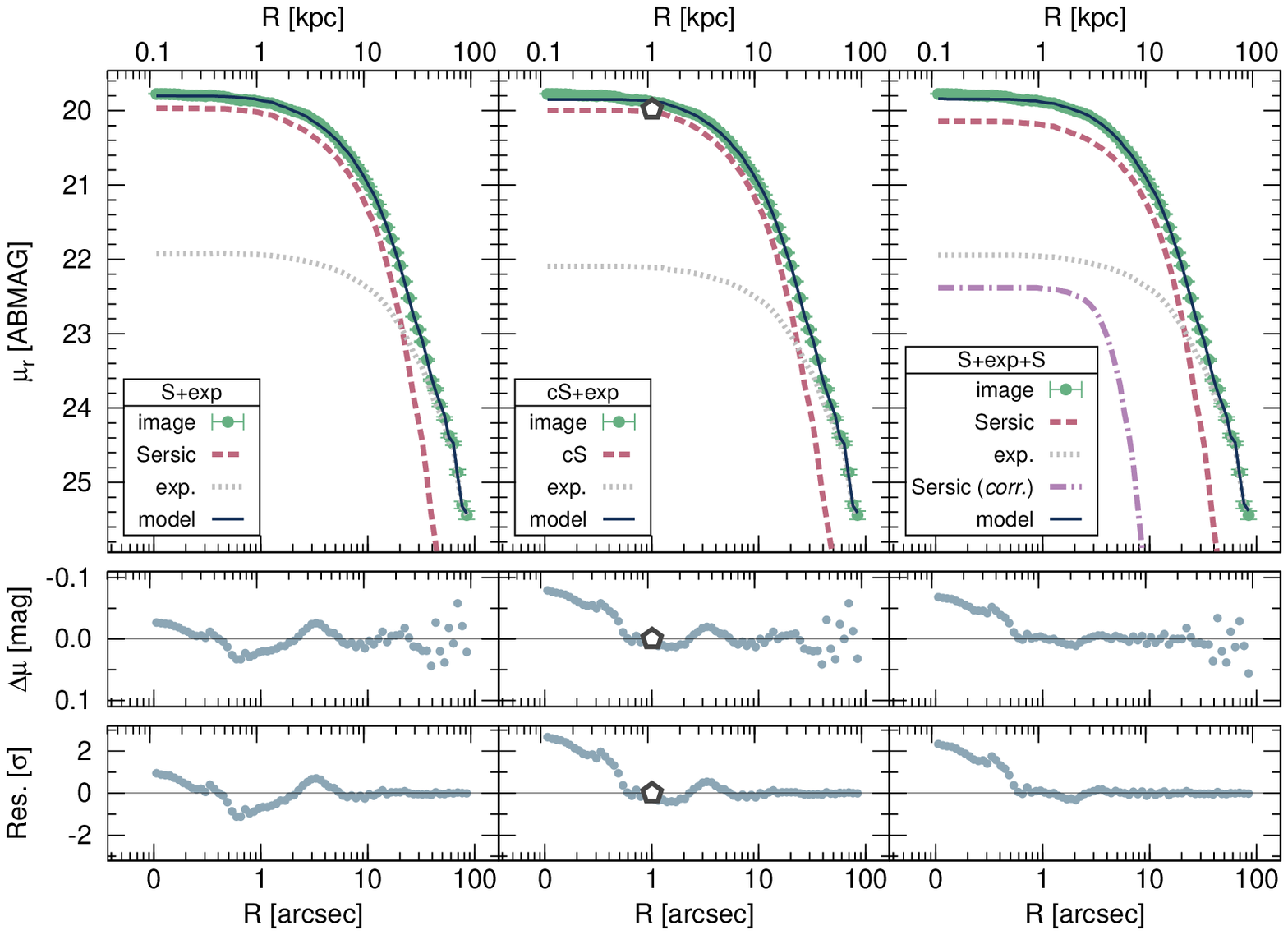}
   \put(5,5){\textcolor{white}{}}
  \end{overpic}

  \\

  \begin{minipage}{0.32\textwidth}
   \begin{overpic}[width=\textwidth]
    {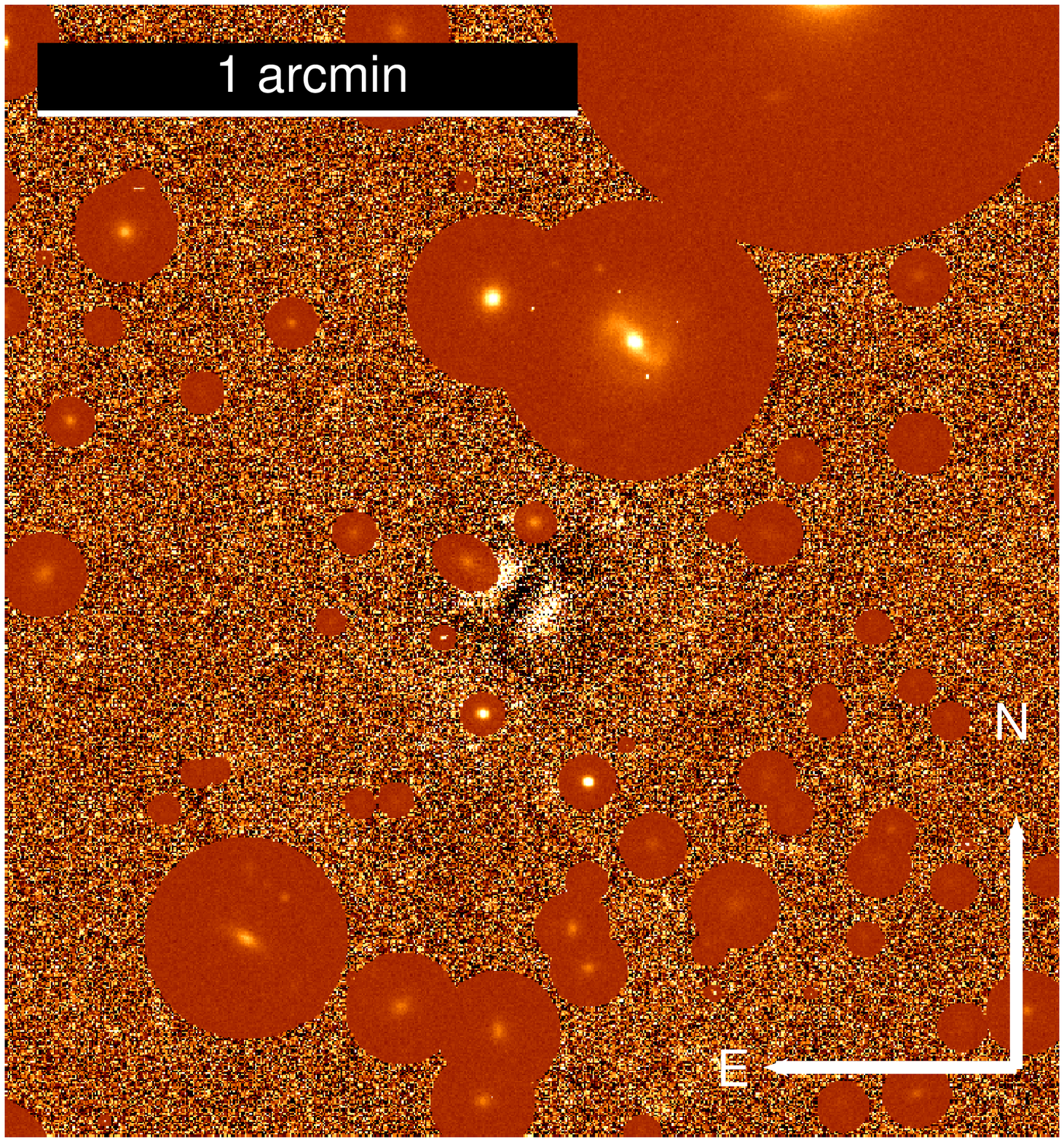}
    \put(5,5){\textcolor{white}{}}
   \end{overpic}
  \end{minipage}

  \begin{minipage}{0.32\textwidth}
   \begin{overpic}[width=\textwidth]
    {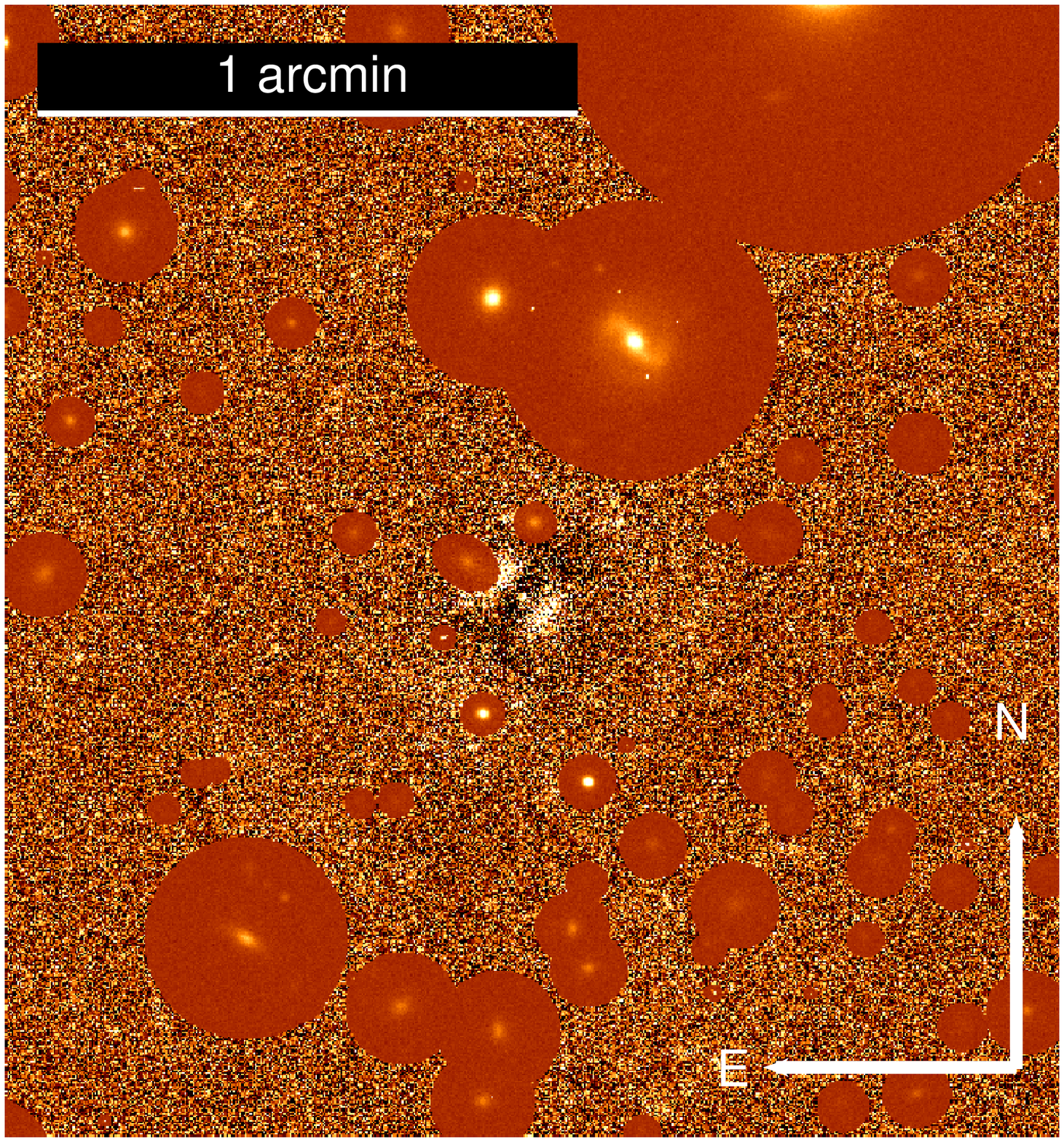}
    \put(5,5){\textcolor{white}{}}
   \end{overpic}
  \end{minipage}
  
  \begin{minipage}{0.32\textwidth}
   \begin{overpic}[width=\textwidth]
    {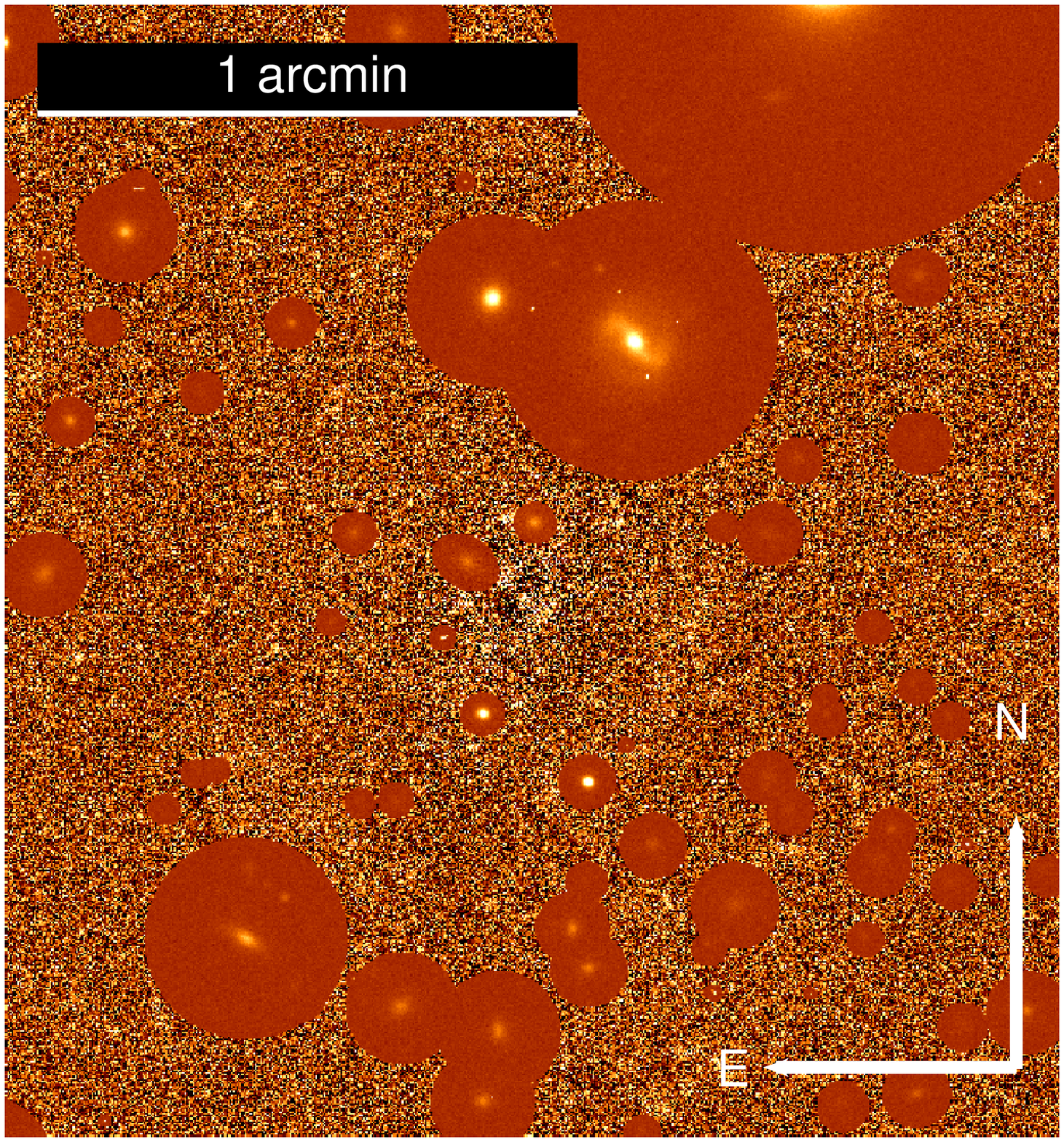}
    \put(5,5){\textcolor{white}{}}
   \end{overpic}
  \end{minipage}

  \\

 \end{tabular}

 \caption{
  2D image analysis: results of the fits to the 2D image of Holm~15A performed assuming:
  a \Sersic{}\-+exponential model (\emph{left}); a \corser{}\-+exponential model   
  (\emph{middle}); and a \Sersic{}\-+exponential model plus an additional inner
  \Sersic{} component intended to compensate for the single ellipticity of the 2D
  model components (\emph{right}).
  \newline
  \emph{Top.-} The green data points represent the major-axis surface brightness
  profile measured over the isophotes defined using IRAF.\emph{ellipse} (i.e. the same
  measurement presented in Figure \ref{figure:ellipse}).
  The curves represent the surface brightness profile of the model images measured over
  exactly the same isophotes.
  The continuous curves show the global models, while the dashed curves represent
  their sub-components.
  We stress that these are \emph{not} fits to the 1D profile, but rather
  surface brightness measurements (projections) of the 2D models.
  The pentagon indicates the location of the \corser{} model's break radius.
  The panels underneath the profiles represent the data residuals about the fitted
  models, first expressed in terms of the difference in surface brightness, and
  then in terms of residuals (in units of counts) divided by the standard deviation
  as measured on the ``sigma'' image.
  \emph{Bottom.-}
  The actual residual images that were minimized by \CORSAIR{}.
  Masked objects are highlighted as in Figure \ref{figure:mosaic_and_mask}.
  \label{figure:fit_2D}
 }
\end{figure*}

\noindent
Overall, our results from the modeling of the 1D light profile and the 2D
image agree very well (see Table \ref{table:fit}).
However, the 1D projection of the 2D residuals (Figure \ref{figure:fit_2D}) show
a more pronounced pattern than seen the 1D residuals (Figure \ref{figure:fit_1D}).
This is partly an artifact of the particular choice of projection, and partly a
direct consequence of each 2D model component of \CORSAIR{} having a single
center and ellipticity weighted over the whole extent of the fit area.
Here we will review the implications of these caveats over two specific regions. 

\smallskip

\noindent
\emph{The innermost $\sim$0$\arcsec$.5}. ---
In the projection shown in Figure \ref{figure:fit_2D}, the 2D model seems to
significantly underestimate the light at radii smaller than $\sim$0$\arcsec$.5.
Within $\sim$0$\arcsec$.5, the isophote centers identified by IRAF.\emph{ellipse}
--- which can trace the centroid of the real galaxy at the sub-pixel level ---
drastically shift (see Figure \ref{figure:ellipse}, bottom panels).
When the photometry is performed over the 2D model along the same isophotes,
the centroid of the \CORSAIR{} model are offset, hence the model brightness is
underestimated within the $\sim$0$\arcsec$.5 region. 

However, since a corresponding tiny ``bump'' is visible in the 1D surface
brightness profile (Figure \ref{figure:ellipse}, top left panel) at those radii,
we investigated whether this small feature should be attributed to an intrinsically
lopsided light distribution, or instead to the presence of a misaligned
point-source.
If there was in fact a point-source, it could be an unresolved nuclear star cluster
or an offset AGN.
Indeed, LC14 found evidence for AGN emission in the optical spectrum of
Holm~15A, which might be visible in our $r$-band image.
We tried to model this potential inner feature by adding a point-source to the
2D models (nominally, a PSF function), but we could not obtain a successful fit
due to the low surface brightness of this potential component relative to the
surrounding galaxy.
We also tried to perform a similar fit in 1D using a PSF and a PSF-convolved
Gaussian (i.e., a slightly extended source), although our 1D residuals
also did not warrant such an additional component.
From this fit we obtained a PSF-convolved Gaussian with FWHM comparable to the
PSF FWHM (as expected for a point-source), and an integrated magnitude for 
any potential nuclear component of $\sim$24.8~mag, which we consider as an upper limit.

\smallskip

\noindent
\emph{The 1$\arcsec$--4$\arcsec$ region}. --- As discussed above, the ellipticity
profile of Holm~15A (Figure \ref{figure:ellipse}, top-right) suggests the presence
of two galaxy components (i.e., spheroid and halo).
Assuming that the 10$\arcsec$--30$\arcsec$ interval marks the transition between
them, our
2D fit seems to recover the ellipticity at the outer edge of each component
(see Table \ref{table:fit}), most probably due to the higher number of pixels along
larger isophotes.
The single ellipticity of the \CORSAIR{} components is the cause for the
pattern visible in the center of the residual image (Figure \ref{figure:fit_2D},
bottom-left), which appears as a 1$\arcsec$-4$\arcsec$ structure oriented in the
direction perpendicular to the galaxian major-axis (i.e., along the NE-SW direction).
This happens because in that region the image isophotes have a much more circular
shape ($\langle$\emph{e}$\rangle$ $\sim$ 0.1) than the 2D \Sersic{} component
(for which we obtained $e$ = 0.2).
In the 1D projection of the 2D residuals (Figure \ref{figure:fit_2D}, top-left), this
pattern manifests as a sinusoidal feature.

\bigskip

\noindent
As a consequence, the 1D and 2D \corser{}\-+\-exponential fits show some differences
in their inner regions.
The 1D model shown in right-hand panels of Figure \ref{figure:fit_1D} yields a small
(sub-pixel) break radius ($R_{b}$ $\sim$ 0$\arcsec$.14), generally consistent with
being a core-\emph{less} galaxy\footnote{
 Recall that the image has a seeing of $\sim$0$\arcsec$.75.
}.
Every \corser{} fit can effectively reproduce an intrinsic \Sersic{}
profile by sufficiently minimizing the break radius of the core.
Although a core is partially resolved in the 2D model (middle panels of Figure \ref{figure:fit_2D}), it does not seem to improve the fit significantly.
In particular, we observe that the 1$\arcsec$--4$\arcsec$ 2D residual feature described
above is still present, and --- although slightly reduced --- still shows the same
pattern, hence reducing the likelihood that it is related to a depleted a core.

On the other hand, outside the problematic inner $\sim$0$\arcsec$.5 region, the
\Sersic{}+exponential and \corser{}+exponential models fit the data comparably well,
yielding similar residuals (see Table \ref{table:fit} and Figures \ref{figure:fit_1D}
and \ref{figure:fit_2D}).
What has occurred is that the \corser{} core parameters were driven by the fit
algorithms to partially compensate for the effects of the $\sim$0$\arcsec$.5 ``excess''
(1D case) or the varying ellipticity (2D case), rather than accounting for an
actual depleted core.

Persuaded that the simpler \Sersic{}\-+exponential model provides the most appropriate
description, we decided to refine the 2D \Sersic{}\-+exponential fit using nested
\CORSAIR{} models to see if we could approximate the varying ellipticity of Holm~15A.
In addition to an ellipticity gradient, the radial range corresponding to the inner
\Sersic{} component is also associated with an irregular position angle profile which
is wildly varying within the innermost 4$\arcsec$ (while it is remarkably constant in
the range where the exponential component dominates; see Figure \ref{figure:ellipse},
middle-right).
Therefore, we paid special attention to improve the fit of the inner component, and we
constructed our refined model by using two \Sersic{} components with different axis
ratios to describe the ``bulge'', plus the previously used exponential function for the ``halo'' (Figure \ref{figure:fit_2D}, right panels).
This extra central component represents a corrective factor, rather than a
distinct element of the galaxy.
The parameters obtained with this new fit are reported in Table \ref{table:fit}.
The best-fit exponential halo component is practically unchanged with the addition of
this new inner \Sersic{} component, indicating that the correction indeed acted mostly
on the ``bulge''.
Moreover, the corrective component has a luminosity one order of magnitude
fainter than the other components, so that the predominant \Sersic{} component
still resembles the previously obtained ``bulge'' (including its position angle
and aspect ratio).
With this refinement the 2D fit drastically improved, now yielding flat residuals
over all the fitted range beyond $\sim$0$\arcsec$.5 (see Figure \ref{figure:fit_2D},
bottom-right panel).
The apparent (false) excess seen here, in the 1D profile is a result of the centroid
shift issue discussed earlier.

\renewcommand{\tabcolsep}{0.7em}

\begin{turnpage}

\begin{deluxetable*}{rccccccccccccccc}
 \renewcommand{\arraystretch}{1.2} 
 \setlength\lightrulewidth{0.01em} 
 \tablecaption{Fit Results\label{table:fit}}
 \tablehead{
  \multicolumn{2}{c}{Model/Component} &
  \colhead{$\mu_{r}^{\dagger}$}       & 
  \colhead{$m_{comp,r}^{\dagger}$}    &
  \colhead{$R_{b}$}                   &
  \colhead{$R_{b}$}                   &
  \colhead{$\alpha$}                  &
  \colhead{$\gamma$}                  &
  \colhead{$R_{e}$}                   &
  \colhead{$R_{e}$}                   &
  \colhead{$n$}                       &
  \colhead{$e$}                       &
  \colhead{P.A.}                      &
  \colhead{$\Delta\mu_{r}$}           &        
  \colhead{$m_{r,0}^{\dagger}$}       &
  \colhead{$M_{r,0}^{\dagger}$}
  \\
  \addlinespace 
  \multicolumn{2}{c}{}         &
  \colhead{[mag/arcsec$^{2}$]} &
  \colhead{[mag]}              &
  \colhead{[kpc]}              &
  \colhead{[arcsec]}           &
  \colhead{}                   &
  \colhead{}                   &
  \colhead{[kpc]}              &
  \colhead{[arcsec]}           &
  \colhead{}                   &
  \colhead{}                   &
  \colhead{[deg]}              &
  \colhead{[mag/arcsec$^{2}$]} &
  \colhead{[mag]}              &
  \colhead{[mag]}
  \\
  \multicolumn{2}{c}{{\tiny (1)}} &
  \colhead{{\tiny (2)}}           &     
  \colhead{{\tiny (3)}}           &
  \colhead{{\tiny (4)}}           &
  \colhead{{\tiny (5)}}           &
  \colhead{{\tiny (6)}}           &
  \colhead{{\tiny (7)}}           &
  \colhead{{\tiny (8)}}           &
  \colhead{{\tiny (9)}}           &
  \colhead{{\tiny (10)}}          &
  \colhead{{\tiny (11)}}          &
  \colhead{{\tiny (12)}}          &
  \colhead{{\tiny (13)}}          &
  \colhead{{\tiny (14)}}          &
  \colhead{{\tiny (15)}}
 }
 \startdata
\multicolumn{14}{c}{1D}\\
 \midrule
 \addlinespace 
\multirow{2}{*}{\Sersic{}+exp. \big\{}           & \Sersic{}                 & 21.60 & 13.78 & $\cdots$ & $\cdots$ & $\cdots$ & $\cdots$ & 13.09  & 11.87  & 1.0   & $\cdots$ & $\cdots$ & \multirow{2}{*}{\big\} 0.015}  & \multirow{2}{*}{12.50} & \multirow{2}{*}{$-$24.52} \\
                                                 & exponential               & 21.93 & 13.32 & $\cdots$ & $\cdots$ & $\cdots$ & $\cdots$ & 46.49  & 42.15  & [1.0] & $\cdots$ & $\cdots$ &                                &                        &                           \\
 \addlinespace 
\multirow{2}{*}{\corser{}+exp. \big\{}           & \corser{}                 & 19.83 & 13.79 & 0.15     & 0.14     & [2.0]    & 0.28     & 12.86  & 11.66  & 0.9   & $\cdots$ & $\cdots$ & \multirow{2}{*}{\big\} 0.014}  & \multirow{2}{*}{12.51} & \multirow{2}{*}{$-$24.49} \\
                                                 & exponential               & 21.94 & 13.33 & $\cdots$ & $\cdots$ & $\cdots$ & $\cdots$ & 46.60  & 42.25  & [1.0] & $\cdots$ & $\cdots$ &                                &                        &                           \\
 \addlinespace 
 \midrule
 \addlinespace 
\multicolumn{14}{c}{2D}\\
 \midrule
 \addlinespace 
\multirow{2}{*}{\Sersic{}+exp. \big\{}            & \Sersic{}                & 21.59 & 13.82 & $\cdots$ & $\cdots$ & $\cdots$ & $\cdots$ & 12.81  & 11.61  & 1.0   & 0.19     & 143.9    & \multirow{2}{*}{\big\} 0.119}  & \multirow{2}{*}{12.61} & \multirow{2}{*}{$-$24.41} \\
                                                  & exponential              & 21.83 & 13.47 & $\cdots$ & $\cdots$ & $\cdots$ & $\cdots$ & 44.08  & 39.96  & [1.0] & 0.38     &	148.0    &                                &	                   &                           \\
 \addlinespace 
\multirow{2}{*}{\corser{}+exp. \big\{}            & \corser{}                & 19.96 & 13.76 & 0.81     & 0.73     & 1.2      & -0.12    & 13.30  & 12.06  & 1.0   & 0.20     & 144.0    & \multirow{2}{*}{\big\} 0.119}  & \multirow{2}{*}{12.62} & \multirow{2}{*}{$-$24.40} \\
                                                  & exponential              & 21.99 & 13.53 & $\cdots$ & $\cdots$ & $\cdots$ & $\cdots$ & 46.90  & 42.52  & [1.0] & 0.40     & 148.2    &                                &                        &                           \\
 \addlinespace 
\multirow{3}{*}{\Sersic{}+exp.+\Sersic{} \Bigg\{} & \Sersic{}                & 21.62 & 13.85 & $\cdots$ & $\cdots$ & $\cdots$ & $\cdots$ & 13.19  & 11.96  & 0.9   & 0.21     & 143.9    & \multirow{3}{*}{\Bigg\} 0.118} & \multirow{3}{*}{12.61} & \multirow{3}{*}{$-$24.41} \\
                                                  & exponential              & 21.85 & 13.46 & $\cdots$ & $\cdots$ & $\cdots$ & $\cdots$ & 44.37  & 40.23  & [1.0] & 0.37     &	148.2    &                                &                        &                           \\
                                                  & \Sersic{} (\emph{corr.}) & 22.67 & 17.82 & $\cdots$ & $\cdots$ & $\cdots$ & $\cdots$ & 4.13   & 3.74   & 0.3   & 0.24     & 54.9     &                                &                        &                           \\
 \addlinespace 
 \enddata
 \tablecomments{
  Best-fit parameters from our 1D and 2D analysis.
  Missing values are not relevant to the model/component under consideration.
  \\
  $^{(1)}$  Fit model/component.
  $^{(2)}$  $r$-band surface brightness at the: break radius (\corser{}), effective radius
            (\Sersic{}), or at $R$ = 0 (exponential).
  $^{(3)}$  Total $r$-band magnitude of the component.
            For the 1D models, we integrated the surface brightness profile assuming
            a constant ellipticity $\langle$e$\rangle$, which we chose as the ellipticity
            at the component's effective radius ($\langle$e$\rangle$ $\sim$ 0.2 for the
            inner components, $\langle$e$\rangle$ $\sim$ 0.3 for the outer exponential
            components; see Figure \ref{figure:ellipse}).
  $^{(4)}$  Break radius in units of kilo-parsecs
            (at the distance of Holm~15A, 1\arcsec = 1.103~kpc).
  $^{(5)}$  Break radius in units of arcseconds.
  $^{(6)}$  Alpha parameter for the \corser{} model; this parameter has been held
            fixed (i.e. not solved for as a free parameter) in the 1D analysis.
  $^{(7)}$  Inner power-law index for the \corser{} model.
  $^{(8)}$  Effective radius ($R_{e}$) of the component in units of kiloparsecs;
            the $R_{e}$ of the \corser{} profile is defined to be the effective
            radius of the whole model in the 1D analysis, while it represents the
            $R_{e}$ of the \Sersic{} part of the composite model in \CORSAIR{};
            since in this case the break radius turns out to be very small, the two
            effectively coincide.
  $^{(9)}$  Effective radius in units of arcseconds.
  $^{(10)}$ \Sersic{} index (for $n$ = 1 the \Sersic{} model corresponds to the
            exponential model).
  $^{(11)}$ Component ellipticity.
  $^{(12)}$ Component position angle (North = 0$^{\circ}$).
  $^{(13)}$ Dispersion of fit residuals.
            For the 1D fits we adopted the RMS of the difference between the light
            profile and the model values (see Figure \ref{figure:fit_1D}).
            For the 2D fits we adopted the inner 68\% ($\pm$1$\sigma$) interval of the
            distribution of values from the residual image pixels (after applying
            the same masking used for the fit).
            Note that the former yields numerically smaller values than the latter
            because the residual data points are averaged over the whole isophote.
  $^{(14)}$ Rest-frame, extinction-corrected $r$-band magnitude of the model.
            A 5log(1+$z$) $\sim$ 0.12~mag magnitude dimming was applied 
            using the distance reported in Table \ref{table:image}.
            Following the prescriptions of \cite{K-corr} for early-type galaxies,
            we also applied an $r$-band $K$-correction $K_{r}$ = 0.06~mag.
            We used the galactic extinction from NED ($\sim$0.09~mag).
  $^{(15)}$ Absolute, rest-frame, extinction-corrected $r$-band magnitude of the
            model (distance modulus is provided in Table \ref{table:image}).
 }
 \tablenotetext{$\dagger$}{
  Values refer to the SDSS $r$-band filter, ABMAG system.
  The zero-point for the calibration is provided in the CADC image header.
 }
\end{deluxetable*}

\end{turnpage}

\section[Discussion]{Discussion}
\label{Discussion}

\noindent
Extended halos around BCGs are commonly observed.
Their presence is not only indicated by the need for adding an ``envelope''
component when fitting models, but also by the twist in the outer isophotes of BCGs,
which has been interpreted as a signature of galaxy accretion/interaction
\citep[e.g.][]{porter,gonzalez}.
This picture is supported by numerical simulations, which showed that BCGs,
being at the centers of large dark matter haloes, underwent a vigorous history of
both minor and major merging \citep[e.g.][]{delucia}, and hence may be surrounded
by a halo of tidally stripped stars \citep[e.g.][]{cooper}.

In fact, in their survey of nearby BCGs, \citet[herafter DM11]{donzelli} had
already reported that a \Sersic{}\-+exponential model fits the light profile
of Holm~15A.
We improve on this by: (\emph{1}) investigating the presence of a
depleted core (DM11 did not perform this test plus their data had a FWHM
$\sim$ 1$\arcsec$--2$\arcsec$ and their fits were limited to a minimum radius of
1.5$\times$FWHM); and (\emph{2}) reaching deeper in limiting surface brightness
($\sim$25.5~\magsb{} compared to their 24.5~\magsb{}).
DM11 thus limited their analysis to $\sim$60$\arcsec$, while our fit extends up to
$\sim$80$\arcsec$, hence providing a better constraint on the extended galaxy halo
light.
Our 1D \Sersic{}\-+exponential fit agrees well with the similar fit by DM11.
Given the luminosity--$n$ relation and the findings in \cite{graham:1996}, it
may be surprising to note that the \Sersic{} index that we derive
for the main ``bulge'' of a bright galaxy like Holm~15A ($M_{V}$ $\sim$ --23.8~mag)
is so small ($n$ $\sim$ 1).
However, this low \Sersic{} index can be compared to the  low \Sersic{} indexes
measured for the inner components of several BCGs (e.g.\ DM11), as well as in a
couple of cD galaxies \citep[NGC~4874 and UGC~9799;][]{seigar}.
These low \Sersic{} indexes might be a result of a dramatic galaxy re-shaping
due to one/some of the processes mentioned in the Introduction.
Although these low-$n$ profiles have rather flat inner regions, we are not dealing
with a profile that displays a clear break and downward deviation at small radii ---
as observed with the traditional partially depleted cores in galaxies with larger
($n$ $\gtrsim$ 3) \Sersic{} indexes.

\begin{figure}
 \begin{center}
  \includegraphics[width=0.5\textwidth,angle=0]{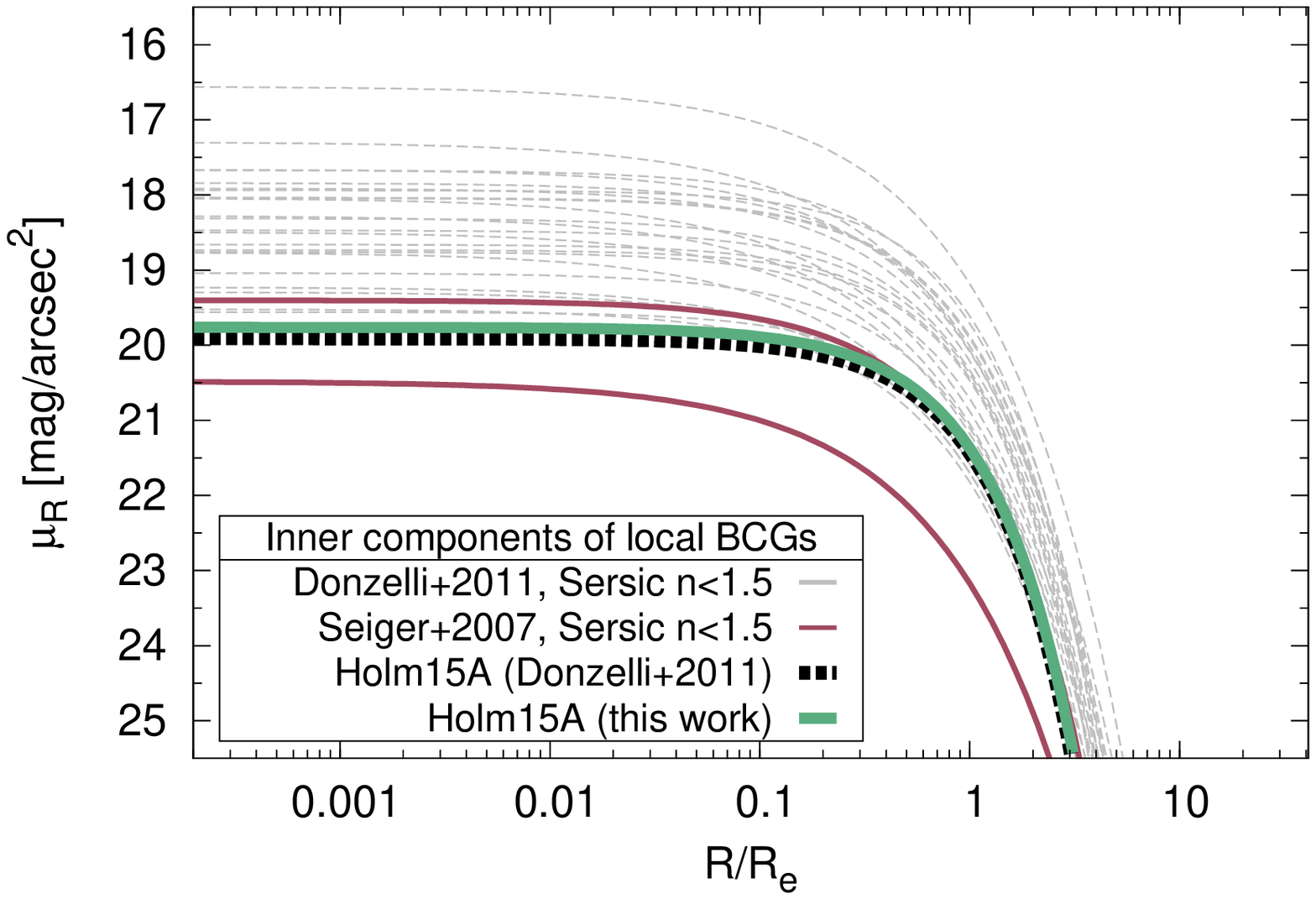}
  \caption{
   Inner components of local BCGs with low \Sersic{} indexes.
   The dashed lines (magenta solid lines) represent the \Sersic{} components of the
   \Sersic{}+exponential fits from \cite{donzelli} \citep{seigar}.
   The black dashed line represents the \cite{donzelli} inner component of Holm~15A,
   while the solid green line shows its spheroidal \Sersic{} component from our
   analysis.
   All the surface brightness profiles are expressed relative to the R-band, and,
   when necessary, have been converted from the observed bands using the colour
   conversions by \cite{fukugita}.
   \label{figure:comparison_BCGs}
  }
 \end{center}
\end{figure}

In Figure \ref{figure:comparison_BCGs} we display the inner galaxy components
from the fits by \cite{donzelli} (their Table 2; dashed lines) and
\cite[][their Table 2; magenta solid lines]{seigar}, after selecting objects
similar to Holm~15A, i.e.\ with \Sersic{}+exponential decomposition, and having
inner spheroids with \Sersic{} $n$ $<$ 1.5.
We highlight the \cite{donzelli} model for Holm~15A with a black dashed line,
which can be observed to closely match the profile we derived (solid green line).
All the surface brightness profiles are expressed relative to the R-band, and,
when necessary, have been converted from the observed bands using the colour
conversions by \cite{fukugita}.
Holm~15A does not seem to show any striking peculiarity in this representation,
except that it has the lowest central surface brightness in the [sub-]sample
from \cite{donzelli}.
This observation strengthens the idea that Holm~15A is an otherwise
somewhat standard representative of the population of local BCGs with low \Sersic{}
index spheroids.

While LC14 favor the Nuker law for their description of the central regions of
Holm~15A, they also use a double-\Sersic{} fit (comparable to our
\Sersic{}\-+exponential model) to recover the total galaxy light, obtaining
\mbox{$m_{V} =$ 13.28~mag}.
We can compare this value with the total magnitude obtained from our 2D
\Sersic{}\-+exponential model, before rest-frame and extinction correction
(\mbox{$m_{r} =$ 12.88~mag}; see Table \ref{table:fit} and table notes).
We converted our magnitude, defined in the SDSS $r$-band (AB system), to the
Johnson/Cousin $V$-band magnitude in the Vegamag system ($m_{V}$) by applying
a \mbox{($V-r'$) = 0.36~mag} color correction, as typical for early-type galaxies
\citep[e.g.][]{fukugita}.
We obtained $m_{V}$ = 13.24~mag, comparable to the measurement of LC14.

An overview on how we came to believe in massive black holes, and the development
of the scaling relations associated with them, can be found in the extensive review
by \cite{graham:review_bh}.
LC14 estimated the mass of the BH in Holm~15A using several scaling relations
(their Table 2), and concluded that the best estimate for \MBH{} is
\mbox{$\sim$10$^{10}$\Msun{}}.
Of the relations they used, the only one which directly connects to the assumed
morphological profile was the \MBH{}--$R_{b}$ relation taken from \cite{rusli},
from which they obtained \MBH{} $\sim$ 1.7$\times$10$^{11}$\Msun{} using their Nuker
law break radius.
This is an order of magnitude above their preferred estimate.
The \cite{rusli} scaling relation was however constructed using the \corser{}
break radius ($R^{cS}_{b}$), which is smaller than the Nuker break radius
($R^{N}_{b}$) for the reasons discussed in the Introduction.
However, we found no convincing evidence in Holm~15A for a depleted core relative
to the inward extrapolation of the outer light profile and as such
we do not deem it appropriate to use the \MBH{}--$R_{b}$ relation.

In the binary black hole scouring scenario, the core mass deficit relates to the
mass of the final merged SMBH \citep[e.g.][]{merritt}. 
The ``depleted'' mass can be inferred from the luminosity deficit with respect
to a model representing the ``unperturbed'' galaxy.
In LC14, they use a de Vaucouleurs profile as their original, unperturbed
model to compare against their double-\Sersic{} fit.
From a visual inspection, their de Vaucouleurs profile
matches the data over the range of galactocentric radii: 10--40 kpc
(LC14; their Figure 2).
However, given that galaxies and BCGs exhibit a variety of \Sersic{} indexes
(e.g.\ DM11), it might not be that the original (i.e., pre-depletion) profile of the
galaxy had $n$ = 4 (corresponding to a de Vaucouleurs profile).
Indeed, the application of a de Vaucouleurs model appears to have produced an
artificial over-sized luminosity deficit.

The discrepancies between our results and those of LC14 are related to the
adopted paradigm for the description of the surface brightness
distribution, i.e.\ the \corser{} or the Nuker framework.
There is no formal mistake in the analysis of LC14, and in fact we were able
to reproduce their Nuker parameters when fitting a Nuker model in 2D with
\GALFIT{}.
However, for the reasons highlighted in the Introduction --- and especially
for its ability to discern real cores from the flat inner slopes of low-$n$
\Sersic{} profiles --- we favor the \corser{} model, which found no evidence
for a depleted core larger than the resolution limit.

Finally, we remark that the depletion of stars from the core of a galaxy due to
coalescing black holes, and other mechanisms, preferentially removes stars on radial
orbits, leaving an excess of stars on tangential orbits \citep[e.g.][]{thomas}.
Integral-field kinematic data may therefore be of benefit to help
identify or reject the presence of a core depleted of stars.

\section[Conclusions]{Conclusions}
\label{Conclusions}

\noindent
We performed a surface brightness analysis of the BCG Holm~15A using a
CFHT-MegaPrime $r$-band image to investigate the claim that this galaxy has the
largest depleted core ever detected \citep[4.57~kpc;][]{lopez}.

We fit the 1D light profile and the 2D image of Holm~15A to compare a
core-\emph{less} galaxy plus envelope model (\Sersic{}\-+exponential) against a
core-\Sersic{} galaxy plus envelope model (\corser{}\-+exponential).
We obtained good agreement among the best-fit parameters derived with the 1D and
2D methods (see Table \ref{table:fit}), modulo some minor differences predominantly
attributable to ellipticity gradients
(\S\ref{Fit comparison and choice of best-fit model}) and a varying center with
isophotal radius.
In order to approximate the varying ellipticity of the inner
galaxy, the 2D models --- having components with fixed ellipticity --- required
the addition of a ``corrective'' component to the \Sersic{}\-+exponential model.

We find that the \corser{} model does not provide an appropriate representation
of the galaxy light distribution.
In particular, in the 1D description the core has a slight excess of light,
while in the 2D model the inner power-law of the \corser{} fit does not represent
an actual real core, but rather compensates for the ellipticity gradient
(\S\ref{Fit comparison and choice of best-fit model}).
We therefore conclude that the galaxy is core-\emph{less} and we favor the idea
that its light distribution is best described by a simple \Sersic{} profile
with a low index $n$ plus an exponential ``halo'' component as included by LC14.

\acknowledgments

The authors wish to thank L. Cortese and P. A. Duc for useful insights on the data
analysis.
We are also very grateful to the authors of LC14 for their availability in
discussing our results. 
This research was supported under the Australian Research
Council's funding scheme (DP110103509 and FT110100263).






\begin{thebibliography}{}

\bibitem[Begelman et al.(1980)]{begelman} Begelman, M.~C., 
Blandford, R.~D., \& Rees, M.~J.\ 1980, \nat, 287, 307 

\bibitem[Bell et al.(2006)]{bell} Bell, E.~F., Phleps, S., 
Somerville, R.~S., et al.\ 2006, \apj, 652, 270 

\bibitem[Bertin \& Arnouts(1996)]{SExtractor} Bertin, E., \& Arnouts, S.\ 1996, \aaps, 117, 393 

\bibitem[Bertin(2011)]{psfex} Bertin, E.\ 2011, Astronomical 
Data Analysis Software and Systems XX, 442, 435

\bibitem[Bonfini(2014)]{corsair} Bonfini, P.\ 2014, \pasp, 126, 935

\bibitem[Boylan-Kolchin et al.(2004)]{boylan} Boylan-Kolchin, 
M., Ma, C.-P., \& Quataert, E.\ 2004, \apjl, 613, L37 

\bibitem[Byun et al.(1996)]{byun:1996} Byun, Y.-I., Grillmair, 
C.~J., Faber, S.~M., et al.\ 1996, \aj, 111, 1889 

\bibitem[Carollo et al.(1997)]{carollo} Carollo, C.~M., Franx, 
M., Illingworth, G.~D., \& Forbes, D.~A.\ 1997, \apj, 481, 710 

\bibitem[Casteels et al.(2014)]{casteels} Casteels, K.~R.~V., 
Conselice, C.~J., Bamford, S.~P., et al.\ 2014, \mnras, 445, 1157

\bibitem[Cooper et al.(2014)]{cooper} Cooper, A.~P., Gao, L., 
Guo, Q., et al.\ 2014, arXiv:1407.5627 

\bibitem[De Lucia 
\& Blaizot(2007)]{delucia} De Lucia, G., \& Blaizot, J.\ 2007, \mnras, 375, 2 

\bibitem[De Propris et al.(2007)]{depropris} De Propris, R., 
Conselice, C.~J., Liske, J., et al.\ 2007, \apj, 666, 212

\bibitem[de Vaucouleurs(1948)]{devaucouleurs} de Vaucouleurs, G.\ 
1948, Annales d'Astrophysique, 11, 247 

\bibitem[Donzelli et al.(2011)]{donzelli} Donzelli, C.~J., 
Muriel, H., \& Madrid, J.~P.\ 2011, \apjs, 195, 15

\bibitem[Duc et al.(2015)]{duc:2014} Duc, P.-A., Cuillandre, J.-C., Karabal, E., et al.\ 2015, \mnras, 446, 120 

\bibitem[Dullo \& Graham(2012)]{dullo:2012} Dullo, B.~T., \& Graham, A.~W.\ 2012, \apj, 755, 163 

\bibitem[Dullo \& Graham(2013)]{dullo:2013} Dullo, B.~T., \& Graham, A.~W.\ 2013, \apj, 768, 36 

\bibitem[Dullo \& Graham(2014)]{dullo:2014} Dullo, B.~T., \& Graham, A.~W.\ 2014, \mnras, 444, 2700

\bibitem[Ebisuzaki et al.(1991)]{ebisuzaki} Ebisuzaki, T., 
Makino, J., \& Okumura, S.~K.\ 1991, \nat, 354, 212 

\bibitem[Faber et al.(1997)]{faber} Faber, S.~M., Tremaine, 
S., Ajhar, E.~A., et al.\ 1997, \aj, 114, 1771 

\bibitem[Ferrarese et al.(2006)]{ferrarese:2006} Ferrarese, L., 
C{\^o}t{\'e}, P., Jord{\'a}n, A., et al.\ 2006, \apjs, 164, 334 

\bibitem[Fukugita et al.(1995)]{fukugita} Fukugita, M., 
Shimasaku, K., \& Ichikawa, T.\ 1995, \pasp, 107, 945 

\bibitem[Goerdt et al.(2010)]{goerdt} Goerdt, T., Moore, B., 
Read, J.~I., \& Stadel, J.\ 2010, \apj, 725, 1707

\bibitem[Gonzalez et al.(2005)]{gonzalez} Gonzalez, A.~H., 
Zabludoff, A.~I., \& Zaritsky, D.\ 2005, \apj, 618, 195 

\bibitem[Graham(1996)]{graham:1996} Graham, A.~W.\ 1996, \apj, 459, 27

\bibitem[Graham et al.(2003)]{graham:corser} Graham, A.~W., Erwin, 
P., Trujillo, I., \& Asensio Ramos, A.\ 2003, \aj, 125, 2951 

\bibitem[Graham \& Guzm{\'a}n(2003)]{graham:2003} Graham, A.~W.,
\& Guzm{\'a}n, R.\ 2003, \aj, 125, 2936

\bibitem[Graham(2004)]{graham:2004} Graham, A.~W.\ 2004, \apjl, 
613, L33 

\bibitem[Graham(2013)]{graham:review_profiles}
Graham, A.W.\ 2013, in "Planets, Stars and Stellar Systems", Volume 6, T.D.Oswalt
\& W.C.Keel  (Eds.), Springer Publishing, p.91 (arXiv:1108.0997)

\bibitem[Graham \& Scott(2013)]{graham:2013} Graham, A.~W., \& Scott, N.\ 2013, \apj, 764, 151 

\bibitem[Graham(2015a)]{graham:review_bh} Graham, A.~W.\ 2015a, arXiv:1501.02937 

\bibitem[Graham(2015b)]{graham:2015}
Graham, A.W.\ 2015b, in Galactic Bulges, E. Laurikainen, R.F. Peletier \& D.Gadotti (Eds.), 
Springer Publishing (arXiv:1501.02937)

\bibitem[Grillmair et al.(1994)]{grillmair:nuker} Grillmair, C.~J., 
Faber, S.~M., Lauer, T.~R., et al.\ 1994, \aj, 108, 102 

\bibitem[Gualandris \& Merritt(2008)]{gualandris} Gualandris, A., \& Merritt, D.\ 2008, \apj, 678, 780 

\bibitem[Jedrzejewski(1987)]{ellipse} Jedrzejewski, R.~I.\ 
1987, \mnras, 226, 747

\bibitem[Hyde et al.(2008)]{hyde} Hyde, J.~B., Bernardi, M., 
Sheth, R.~K., \& Nichol, R.~C.\ 2008, \mnras, 391, 1559 

\bibitem[Laporte \& White(2014)]{laporte} Laporte, C.~F.~P., \& White, S.~D.~M.\ 2014, arXiv:1409.1924

\bibitem[Lauer(1983)]{lauer:1983} Lauer, T.~R.\ 1983, in Elliptical Galaxies,
Surface Photometry, Santa Cruz: University of California

\bibitem[Lauer et al.(1995)]{lauer:nuker} Lauer, T.~R., Ajhar, 
E.~A., Byun, Y.-I., et al.\ 1995, \aj, 110, 2622

\bibitem[Lauer et al.(2007)]{lauer:2007} Lauer, T.~R., Faber, 
S.~M., Richstone, D., et al.\ 2007, \apj, 662, 808 

\bibitem[L{\'o}pez-Cruz et al.(2014)]{lopez} L{\'o}pez-Cruz, 
O., A{\~n}orve, C., Birkinshaw, M., et al.\ 2014, \apjl, 795, LL31

\bibitem[Milosavljevi{\'c} \& Merritt(2001)]{milosavljevic} Milosavljevi{\'c}, M., \& Merritt, D.\ 2001, \apj, 563, 34 

\bibitem[King(1962)]{king:model_1} King, I.\ 1962, \aj, 67, 471

\bibitem[King(1966)]{king:model_2} King, I.~R.\ 1966, \aj, 71, 64 

\bibitem[King \& Minkowski(1966)]{king:1966} King, I.~R., \& Minkowski, R.\ 1966, \apj, 143, 1002 

\bibitem[King \& Minkowski(1972)]{king:1972} King, I.~R., \& Minkowski, R.\ 1972, External Galaxies and Quasi-Stellar Objects, 44, 87 

\bibitem[Kormendy(1982)]{kormendy:1982} Kormendy, J.\ 1982, Saas-Fee 
Advanced Course 12: Morphology and Dynamics of Galaxies, 113

\bibitem[Kormendy et al.(1994)]{kormendy:1994} Kormendy, J., 
Dressler, A., Byun, Y.~I., et al.\ 1994, European Southern Observatory 
Conference and Workshop Proceedings, 49, 147

\bibitem[Kormendy 
\& Bender(2009)]{kormendy:2009} Kormendy, J., \& Bender, R.\ 2009, \apjl, 691, L142 

\bibitem[Kormendy \& Ho(2013)]{kormendy:review} Kormendy, J., \& Ho, L.~C.\ 2013, \araa, 51, 511

\bibitem[Krajnovi{\'c} et al.(2006)]{krajnovic:2006} Krajnovi{\'c}, 
D., Cappellari, M., de Zeeuw, P.~T., \& Copin, Y.\ 2006, \mnras, 366, 787 

\bibitem[Krajnovi{\'c} et al.(2013)]{krajnovic:2013} Krajnovi{\'c}, 
D., Karick, A.~M., Davies, R.~L., et al.\ 2013, \mnras, 433, 2812

\bibitem[\protect\citeauthoryear{Kulkarni \& Loeb}{2012}]{kulkarni} Kulkarni G., Loeb A., 2012, MNRAS, 422, 1306 

\bibitem[Magnier 
\& Cuillandre(2004)]{elixir} Magnier, E.~A., \& Cuillandre, J.-C.\ 2004, \pasp, 116, 449 

\bibitem[Martizzi et al.(2012)]{martizzi} Martizzi, D., Teyssier, R., \& Moore, B.\ 2012, \mnras, 420, 2859 

\bibitem[Merritt et al.(2004)]{merritt:2004} Merritt, D., 
Milosavljevi{\'c}, M., Favata, M., Hughes, S.~A., 
\& Holz, D.~E.\ 2004, \apjl, 607, L9 

\bibitem[Merritt(2006)]{merritt} Merritt, D.\ 2006, \apj, 648, 976

\bibitem[Peng et al.(2010)]{GALFIT} Peng, C.~Y., Ho, L.~C., 
Impey, C.~D., \& Rix, H.-W.\ 2010, \aj, 139, 2097

\bibitem[Pierini et al.(2008)]{pierini} Pierini, D., Zibetti, S., Braglia, F., et al.\ 2008, \aap, 483, 727

\bibitem[Planck Collaboration et al.(2014)]{planck} Planck Collaboration, Ade, P.~A.~R., Aghanim, N., et al.\ 2014, \aap, 571, AA16 

\bibitem[Poggianti(1997)]{K-corr} Poggianti, B.~M.\ 1997, \aaps, 122, 399 

\bibitem[Porter et al.(1991)]{porter} Porter, A.~C., 
Schneider, D.~P., \& Hoessel, J.~G.\ 1991, \aj, 101, 1561

\bibitem[Postman et al.(2012)]{postman} Postman, M., Lauer, 
T.~R., Donahue, M., et al.\ 2012, \apj, 756, 159 

\bibitem[Ravindranath et al.(2001)]{ravindranath} Ravindranath, S., 
Ho, L.~C., Peng, C.~Y., Filippenko, A.~V., 
\& Sargent, W.~L.~W.\ 2001, \aj, 122, 653

\bibitem[Rest et al.(2001)]{rest} Rest, A., van den Bosch, 
F.~C., Jaffe, W., et al.\ 2001, \aj, 121, 2431

\bibitem[Rusli et al.(2013)]{rusli} Rusli, S.~P., Erwin, P., 
Saglia, R.~P., et al.\ 2013, \aj, 146, 160

\bibitem[Redmount \& Rees(1989)]{redmount} Redmount, I.~H., \& Rees, M.~J.\ 1989, Comments on Astrophysics, 14, 165

\bibitem[Richings et al.(2011)]{richings} Richings, A.~J., Uttley, P., K\"{o}rding, E.\ 2011, \mnras, 415, 2158 

\bibitem[Savorgnan \& Graham(2015)]{savorgnan:2015} Savorgnan, G.~A.~D.,
\& Graham, A.~W.\ 2015, \mnras, 446, 2330 

\bibitem[Seigar et al.(2007)]{seigar} Seigar, M.~S., Graham, 
A.~W., \& Jerjen, H.\ 2007, \mnras, 378, 1575 

\bibitem[Thomas et al.(2014)]{thomas} Thomas, J., Saglia, R.~P., Bender, R., Erwin, P., \& Fabricius, M.\ 2014, \apj, 782, 39 

\bibitem[Trujillo et al.(2004)]{trujillo:corser} Trujillo, I., Erwin, 
P., Asensio Ramos, A., \& Graham, A.~W.\ 2004, \aj, 127, 1917 

\bibitem[Volonteri \& Ciotti(2013)]{volontieri} Volonteri, M., \& Ciotti, L.\ 2013, \apj, 768, 29


\end{thebibliography}
\end{document}